\documentclass{article}
\usepackage{latexsym,e-journal}
\usepackage{amssymb,amsfonts}       
\usepackage{times}
\usepackage{epic}
\usepackage{epsfig}
\usepackage{subfigure}
\usepackage{amsmath}
\usepackage{rotating}
\usepackage{curves}
\begin{document}

\newcommand{\rum}{\rule{0.5pt}{0pt}}
\newcommand{\rub}{\rule{1pt}{0pt}}
\newcommand{\rim}{\rule{0.3pt}{0pt}}
\newcommand{\numtimes}{\mbox{\raisebox{1.5pt}{${\scriptscriptstyle \times}$}}}
\newcommand{\numtimess}{\mbox{\raisebox{1pt}{${\scriptscriptstyle \times}$}}}
\newcommand{\Boldsq}{\vbox{\hrule height 0.7pt
\hbox{\vrule width 0.7pt \phantom{\footnotesize T}%
\vrule width 0.7pt}\hrule height 0.7pt}}

\renewcommand{\refname}{References}
\renewcommand{\figurename}{\small Fig.}

\newcommand{\abs}[1]{\mid#1\mid}
\newcommand{\DoT}[1]{\begin{turn}{-180}\raisebox{-1.7ex}{#1}\end{turn}}
\newcommand{\YD}{%
\put(0,0){\curve(3.3,2.0, 4.0,0.3, 3.3,-0.3, 2.8,0.3, 3.3,2.0)}
\DoT{\DoT{\sf Y}}}
\newcommand{\YL}{%
\setlength{\unitlength}{0.1em}
\put(0,0){\curve(2.6,4.0, 1.2,5.3, 0.4,6.6, 0.5,7.4, 1.4,7.3, 1.9,5.8, 2.6,4.0)}
\DoT{\DoT{\sf Y}}}
\newcommand{\YR}{%
\setlength{\unitlength}{0.1em}
\put(0,0){\curve(4.2,4.0, 4.8,6.1, 5.2,6.9, 6.4,7.2, 6.3,6.6, 5.8,5.9, 4.2,4.0)}
\DoT{\DoT{\sf Y}}}
\newcommand{\TYL}{\DoT{\YL}}
\newcommand{\TYR}{\DoT{\YR}}
\newcommand{\TYD}{\DoT{\YD}}
\newcommand{\oYR}{\DoT{\DoT{$\overline{\YR}$}}}
\newcommand{\oYL}{\DoT{\DoT{$\overline{\YL}$}}}
\newcommand{\oYD}{\DoT{\DoT{$\overline{\YD}$}}}
\newcommand{\oTYL}{\DoT{\DoT{$\overline{\TYL}$}}}
\newcommand{\oTYR}{\DoT{\DoT{$\overline{\TYR}$}}}
\newcommand{\oTYD}{\DoT{\DoT{$\overline{\TYD}$}}}
\newcommand{\Y}{\DoT{\DoT{{\sf Y}}}}

\twocolumn[%
\begin{center}
{\Large\bf Fermions as Topological Objects\rule{0pt}{0pt}}\par
\bigskip
Vladimir N. Yershov \\ 
{\small\it  Mullard Space Science Laboratory 
(University College London)\rule{0pt}{13pt}},\\ 
\raisebox{1pt}{\small\it  Holmbury St. Mary
(Dorking), Surrey RH5 6NT, England}\\
\raisebox{-1pt}{\footnotesize E-mail: vny@mssl.ucl.ac.uk}\par
\bigskip\smallskip
{\small\parbox{11cm}{%
A preon-based composite model of the fundamental fermions is discussed, in which 
the fermions are bound states of smaller entities --- primitive charges (preons).
The preon is regarded as a dislocation in a dual 3-dimensional
manifold --- a topological object with no properties, save its unit mass and unit charge. 
It is shown that the dualism of this manifold gives rise to a hierarchy of complex 
structures resembling by their properties three families of the fundamental
fermions. Although just a scheme for building a model of elementary particles,
this description yields a quantitative explanation of many observable particle
properties, including their masses. PACS numbers: 12.60.Rc, 12.15.Ff, 12.10.Dm
\rule[-6pt]{0pt}{0pt}}}\medskip
\end{center}]{%

\setcounter{section}{0}
\setcounter{figure}{0}
\setcounter{table}{0}
\setcounter{equation}{0}

\label{art-44}

\markboth{V.\,N.\,Yershov. Fermions as Topological Objects}{\thepage}
\markright{V.\,N.\,Yershov. Fermions as Topological Objects}

\section{Introduction}

The hierarchical pattern observed in the properties 
of the fundamental fermions (quarks and leptons) 
points to their composite nature \cite{Kalman05},
which goes beyond the scope of the Standard Model of particle
physics. The particles are group\-ed into
three generations (families), each containing two quarks and two leptons with their
electric charges, spins and other properties repeating from generation 
to generation: the electron and its
neutrino, $e^-, \nu_e$, the muon and its neutrino,  $\mu^-,\nu_\mu$,
the tau and its neutrino, $\tau^-,\nu_\tau$, 
the up and down quarks, $u^{+2/3}, d^{-1/3}$,
charm and strange, $c^{+2/3}, s^{-1/3}$, 
top and bottom, $t^{+2/3}, b^{-1/3}$
(here the charges of quarks are indicated by superscripts). 
The composite models of quarks and leptons 
\cite{dugne97}
are based on fewer fundamental particles than the Standard Model 
(usually two or three) and are able to reproduce the above pattern as to the
electric and colour charges, spins and, in some cases, 
the variety of species. 
However, the masses of the fundamental fermions
are distributed in a rather odd way \cite{properties}. They cannot 
be pre\-dicted from any application of first principles of the 
Standard Model; nor has any analysis of the observed data 
\cite{pesteil91} or development of new mathematical ideas
\cite{golfand71} yielded an ex\-planation as to why they should 
have strictly the observed values instead of any others. 
Even there exist claims of ran\-domness of this pattern \cite{donoghue98}.
However, the history of science shows that, whenever a regular pattern 
was observed  in the properties of matter (e.g., the periodical table of
elements or eight-fold pattern of mesons and baryons), this pattern could be explained 
by invoking some underlying structures. 
In this paper we shall follow this lead by assuming
that quarks and leptons are bound states of smaller particles,
which are usually called ``pre-quarks'' 
or ``preons'' \cite{souza92}. 
Firstly, we shall guess at the basic symmetries 
of space, suggesting that space, as any other physical 
entity, is dual. We propose that it is this property that 
is responsible for the emergence of different types of interactions
from a unique fundamental interaction.
To be absolutely clear, we have to emphasise that 
our approach will be based on classical (deterministic) fields, 
which is opposed to the commonly-held view that quarks and 
leptons are quantum objects. But we shall see that by using classical 
fields on small scales we can avoid the problems related to 
the short-range energy divergences and anomalies, which is the main
problem of all quantum field theories. 

\markright{V.\,N.\,Yershov. Fermions as Topological Objects}

\vspace*{-1pt}
\section{The universe}

\markright{V.\,N.\,Yershov. Fermions as Topological Objects}

\vspace*{-1pt}
Let us begin from a few conjectures (postulates) about the 
basic properties of space: 
 
\begin{itemize} 
\vspace*{-1pt}
\item [P1] {Matter is structured, and the number of its structural levels
is finite;}
\item [P2] {The simplest (and, at the same time, 
the most complex) structure in the universe is the universe itself;}
\item [P3] {The universe is self-contained (by definition);}
\item [P4] {All objects in the universe spin (including
the universe itself).} 
\end{itemize}

\vspace*{-1pt}\noindent
The postulate P1 is based on the above mentioned historical experience 
with the patterns and structures behind them. 
These patterns are known to be simpler on lower structural 
levels, which suggests that matter 
could be structured down to the simplest possible entity with almost 
no properties.
We shall relate this entity to the structure of the entire 
universe (postulate P2). This is not, of course, a novelty, since considering
the universe as a simple uniform object lies in the heart of modern 
cosmology. The shape (topology) of this object is not derivable 
from Einstein's equations, but for simplicity it is usually 
considered as a hyper-sphere ($S^3$) of positive, negative or zero curvature.
However, taking into account the definition of the universe as a 
self-contained object (postulate P3), the spherical shape becomes 
in\-ap\-prop\-riate,\rule[-6pt]{0pt}{0pt}
because any sphere has at least two unrelated 
hyper-surfaces, which is incompatible with 
the definition of the uniqueness and self-containedness of the universe.
More convenient would be a manifold with a unique hyper-surface, 
such as the Klein-bottle, $K^3$ \cite{kastrup87}. Similarly to $S^3$, it
can be of positive, negative or zero curvature. An important feature of $K^3$
is the unification of its inner and outer surfaces (Fig.~\ref{fig:klein1}).
In the case of the universe, the unification might well occur on the 
sub-quark level, giving rise to the structures of elementary particles 
and, supposedly, resulting in the identification of the global cosmological
scale with the local microscopic scale of el\-ement\-ary particles. 
In Fig.~\ref{fig:klein1}b the unification region is marked  as  $\Pi $ 
(primitive particle).

\begin{figure}[htb]
\vspace{-0.3cm}
\begin{center}
\raisebox{10pt}{\includegraphics[scale=0.19]{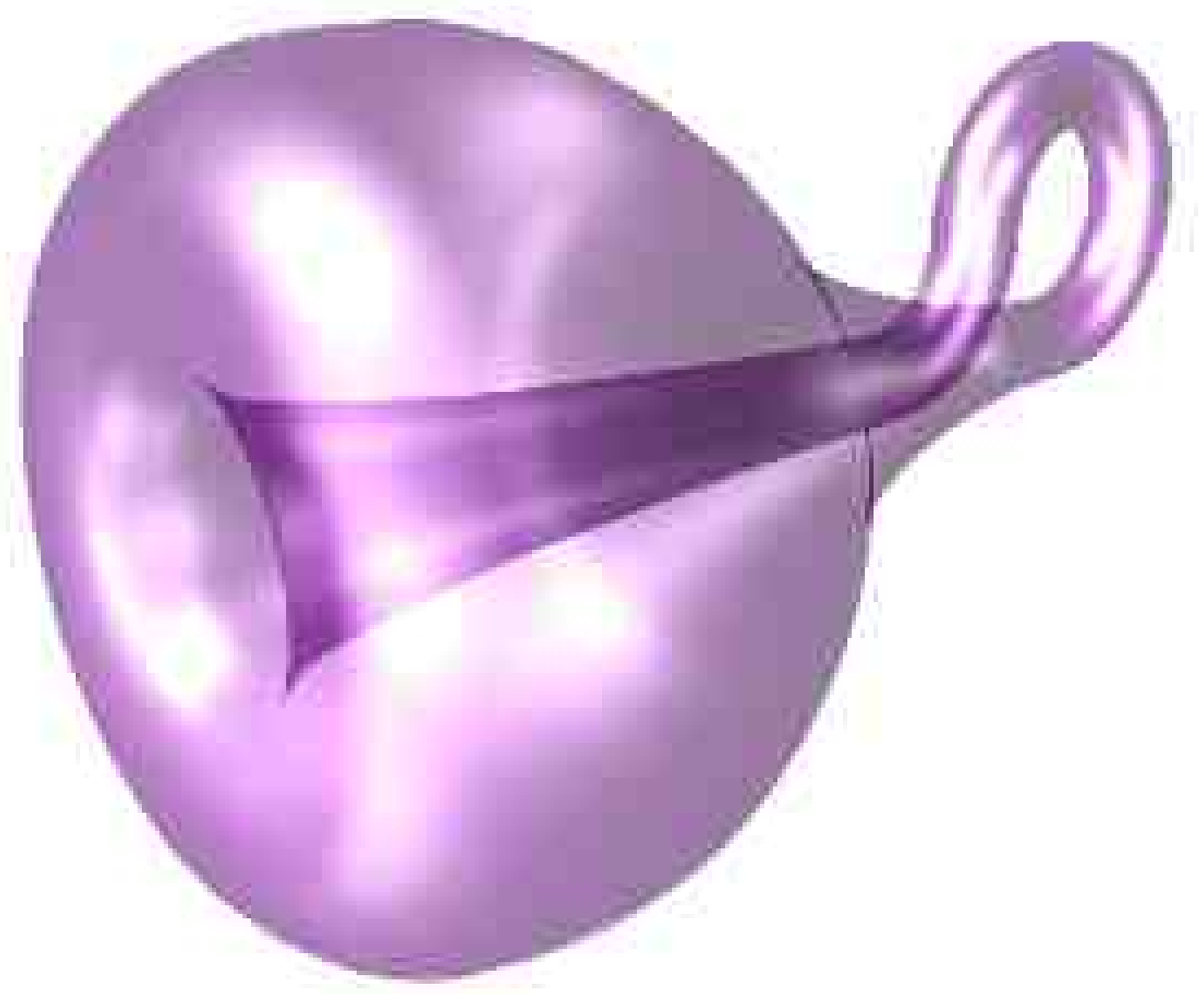}}
\includegraphics[scale=0.9]{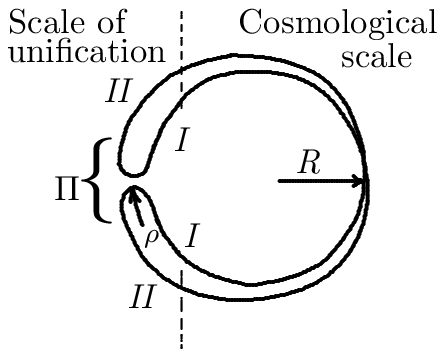}
\put(-190,5){\scriptsize {\bf {\sl a}}}
\put(-53,5){\scriptsize {\bf {\sl b}}}
\vspace{-0.2cm}
\caption{ 
(a) Klein-bottle and (b) its one-dimensional representation;
the ``inner'' (I) and  ``outer'' (II) hyper-surfaces
are unified through the region $\Pi$ (primitive particle);
$R$ and $\rho$ are, respectively, the global and local radii of curvature.
 \label{fig:klein1}
}
\end{center}
\end{figure}
\vspace*{-6mm}

\markright{V.\,N.\,Yershov. Fermions as Topological Objects}

\section{The primitive particle}

\markright{V.\,N.\,Yershov. Fermions as Topological Objects}

Let as assume that space is smooth and continuous, i.e., that 
its local curvature cannot exceed some finite value $\varepsilon$:
$\abs{\rho}^{-1} < \varepsilon$. Then, within the region $\Pi$ (Fig.~\ref{fig:klein1}b)
space will be locally curved  ``inside-out''. In these terms, the primitive particle 
can be seen as a dislocation (topological defect) of the medium and, thus, cannot 
exist independently of this medium. Then, the postulate P4 about the spinning universe 
gives us an insight into the possible origin of the particle mass.   
This postulate is not obvious, although the idea of spinning universe was proposed 
many years ago by A.~Zelmanov \cite{zelmanov44} and K.~G\"odel \cite{godel59}.
It comes from the common fact that so far non-rotating 
objects have never been observed. 

The universe spinning with its angular velocity $\omega$ (of course, if 
considered from the embedding space) would result in the linear 
velocity  ${\pm}\, \omega R$ of the medium in the vicinity of the 
primitive particle, where $R$ is the global radius of curvature of 
the universe; and the sign depends on the choice of the referent 
direction (either inflow or outflow from the inversion region).

Due to the local curvature, ${\rho}^{-1}$, in the vicinity of the 
primitive particle, the spinning  universe must give rise to a local 
acceleration, $a_{\rm g}$, of the medium moving through the region $\Pi$,
which is equivalent to the acceleration of the particle itself. 
According to Newton's second law, this acceleration can be described in terms of 
a force, $F_{\rm g}\,{=}$ 
${=}\,m_{\rm g} a_{\rm g}$, proportional to this acceleration.
The coefficient of pro\-portionality between the acceleration and the
force cor\-res\-ponds to the inertial mass of the particle.
However, for an observer in the coordinate frame of the primitive particle
this mass will be perceived as gravitational ($m_{\rm g}$) because 
the primitive particle is at rest in this coordinate frame. Thus, the spinning
universe implies the accelerated motion of the primitive particle 
along its world line (time-axis).   
If now the particle is forced to move along the spatial
coordinates with an additional acceleration $a_{\rm i}$, it will resist this
acceleration in exactly the same way as it does when accelerating along the
time-axis. A force $F_{\rm i}\,{=}\,m_{\rm i} a_{\rm i}$, which is required in
order to accelerate the particle, is proportional to $a_{\rm i}$ with the
coefficient of proportionality $m_{\rm i}$ (inertial mass). 
But, actually, we can see that within our framework
the inertial, $m_{\rm i}$, and gravitational, $m_{\rm g}$, masses are generated
by the same mechanism of acceleration. That is, mass in this framework is a purely 
inertial phenomenon ($m_{\rm i}\equiv m_{\rm g}$).  

It is seen that changing the sign of $\omega R$ 
does not change the sign of the second derivative 
$a_{\rm g}=\frac{\partial^2 (ict)}{\partial t^2\rule{0pt}{6.0pt}}$, i.\,e.,
of the ``gravitational'' force $F_{\rm g}=m_{\rm g}a_{\rm g}$.
This is obvious, because the local curvature, ${\rho}^{-1}$, is the property
of the manifold and does not depend on the direction of motion. 
By contrast, the first derivative $\frac{\partial (ict)}{\partial t\rule{0pt}{6.0pt}}$
can be either positive or negative, depending on the choice of the 
referent direction. It would be natural here to identify the 
corresponding force as electro\-static. 
For simplicity, in this paper we shall use unit values for 
the mass and electric charge of the primitive
particle, denoting them as $m_\circ$ and $q_\circ$.

In fact, the above mass acquisition scheme has to be
mod\-ified because, besides the local curvature, one must 
account for torsion \rule{-0.6pt}{0pt}of the manifold (corresponding to the
\rule{-0.9pt}{0pt}Weyl \rule{-0.6pt}{0pt}tensor). 
In the three-dimensional case, torsion has three degrees of freedom,
and the corresponding field can be resolved into three components
(six --- when both manifestations of space, $I$ and $II$, are taken into account).
It is reasonable to relate these three components 
to three polarities (colours) of the strong interaction. 
 
Given two manifestations of space, 
we can resolve the field of the particle into two components,
$ \phi_{\rm s}$ and $\phi_{\rm e}$. To avoid singularities we shall
assume that infinite energies are not accessible
in nature. Then, since it is an experimental fact that energy usually increases
as distance decreases, we can hypothesise that the energy of both 
$\phi_{\rm e}$ and $\phi_{\rm s}$, after reaching a maximum, decays to zero
at the origin. The simplest form for the split field that incorporates
the requirements above is the following:
\vspace{-0.1cm}
\begin{equation}
\begin{split}
F&=\phi_{\rm s}+\phi_{\rm e}\,, \\
\phi_{\rm s}&=s \exp(-\rho^{-1})\,, \hspace{0.3cm} \phi_{\rm e}=-\phi'_{\rm s}(\rho)\,.
\end{split}
\label{eq:f2}
\end{equation}

\vspace*{-0.1cm}
Here the signature $s\,{=}\,{\pm}\rub 1$ indicates the sense of the interaction
(attraction or repulsion); the derivative of $\phi_{\rm s}$ is taken with
respect to the radial coordinate $\rho$. Far from the source, the second
component of the split field $F$ mimics the Coulomb gauge, whereas the 
first component extends to infinity being almost constant (similarly to the
strong field). 
 
In order to formalise the use of tripolar fields we have to
introduce a set of auxiliary $3 \times 3$ singular matrices $\Pi^i$ with the 
following elements:
\begin{equation}
^\pm \pi^i_{jk}=\pm \delta^i_j(-1)^{\delta^k_j},
\end{equation}
where $\delta^i_j$ is the Kronecker delta-function; the ($\pm$)-signs cor\-res\-pond
to the sign of the charge; and the index $i$ stands for the colour
($i=1,2,3$ or red, green and blue). The diverging components of the field
can be represented by reciprocal elements: 
%
$\tilde{\pi}_{jk}=\pi_{jk}^{-1}.$
%
Then we can define the (unit) charges and masses of the primitive particles
by summation of these matrix elements:
\begin{equation}
\begin{split}
q_\Pi&=\mathbf{u}^\intercal \Pi \mathbf{u}, \hspace{1.5cm} 
\tilde{q}_\Pi=\mathbf{u}^\intercal \tilde{\Pi} \mathbf{u} \\
m_\Pi&=\abs{\mathbf{u}^\intercal \Pi \mathbf{u}}, \hspace{1cm}
\tilde{m}_\Pi=\abs{\mathbf{u}^\intercal \tilde{\Pi} \mathbf{u}}
\end{split}
\label{eq:superpositionmass}
\end{equation}
($\mathbf{u}$ is the diagonal of a unit matrix; $\tilde{q}_\Pi$ and
$\tilde{m}_\Pi$ diverge).
As\-sum\-ing that the strong and electric interactions are
mani\-fest\-at\-ions of the same entity 
and taking into account the known pattern \cite{suisso02} of the colour-interaction
(two like-charged but unlike-coloured particles are attracted, otherwise they repel),
we can write the signature $s_{ij}$ of the chromoelectric inter\-action
between two primitive particles, say of the colours $i$ and $j$, as:
\begin{equation}
s_{ij}=-\mathbf{u}^\intercal \Pi^i \rub\Pi^j \mathbf{u}\,.
\label{eq:seforce}
\end{equation}

\markright{V.\,N.\,Yershov. Fermions as Topological Objects}

\section{Colour dipoles}

\markright{V.\,N.\,Yershov. Fermions as Topological Objects}

Obviously, the simplest structures allowed by the tripolar field are 
the monopoles, dipoles and tripoles, unlike the con\-ventional bipolar (electric)
field, which allows only the mono\-poles and dipoles.
Let us first consider the colour-dipole configuration. It follows from 
(\ref{eq:seforce}) that two like-charged part\-icles with unlike-colours will
combine and form a charged colour-dipole, $g ^\pm$.
Similarly, a neutral colour-dipole, $g^0$, can also be formed
--- when the constituents of the dipole have unlike-charges. 

\begin{figure}[htb]
\vspace{-0.5cm}
\centering
\begin{turn}{-90}
\includegraphics[scale=1.0]{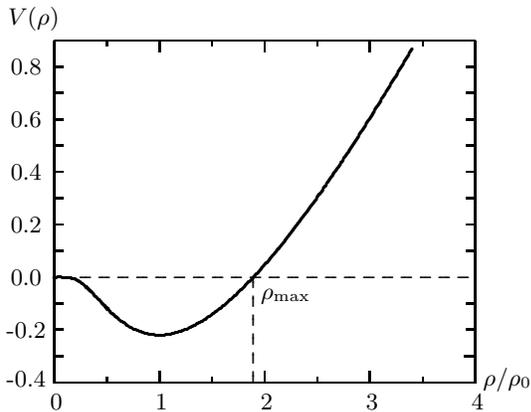}
\end{turn}

\vspace*{-7mm}
\caption{Equilibrium potential based on the split field 
(\ref{eq:f2})
\label{fig:potential}
}
\vspace*{-4mm}
\end{figure}

The dipoles $g^\pm$ and $g^0$ are classical oscillators with the 
double-well potential $V(\rho)$, Fig.~\ref{fig:potential}, derived 
from the split field (\ref{eq:f2}). The oscillations take place 
within the region\linebreak $\rho \in \!(0,\rho_{\rm \,max})$, with the 
maximal distance between the com\-po\-nents $\rho_{\rm max} \,{\approx }\,1.894 \rho_\circ$
(assuming the initial condition $E_0\,{=}$ ${=}\,V(0)$ and setting this energy to zero).
 
Let us assume that the field $F(\rho)$ does not act 
instant\-an\-eously at a distance. Then, we can define the mass of a system with, say, $N$
primitive particles as proportional to the number of these particles, wherever
the field flow rate is not cancelled. For this purpose we shall regard 
the total field flow rate, $v_N$, of such a system as a superposition 
of the individual volume flow rates of its $N$ constituents. Then the 
net mass of the system can be calculated (to a first-order of approximation)
as the number of particles, $N$, times the normalised to unity (Lorentz-additive)
field flow rate $v_N^{\phantom{0}}$:
\begin{equation}
m_N^{\phantom{0}}=\abs N v_N^{\phantom{0}}.
\label{eq:mass}
\end{equation}

Here $v_N^{\phantom{0}}$ is calculated recursively from
\begin{equation} 
v_i=\frac{q_i+v_{i-1}}{1+\abs q_i v_{i-1}},
\label{eq:mass2}
\end{equation}
with $i\,{=}\,2, \dots , N$ and putting $v_1\,{=}\,q_1$.  
Then, when two\linebreak unlike-charged particles combine 
(say red and antigreen), the magnitudes of their oppositely directed 
flow rates cancel each other (resulting in a neutral system). 
The corresponding, acceleration also vanishes, 
which is implicit in (\ref{eq:mass}), 
formal\-is\-ing the fact that the mass of a neutral system is null\-i\-fied.
This formula implies the complete cancellation of
masses in the systems with vanishing electric fields, but this 
is only an approximation because in our case  
the primitive particles are separated by the average distance $\rho_\circ$,
whereas the complete cancellation of flows is possible only when the
flow source centres coincide.

In the matrix notation, the positively charged 
dipole, $g^+_{12}$, is represented as a sum of two matrices, 
$\Pi^1$ and $\Pi^2$:
\begin{equation}
g^+_{12} =\Pi^1+\Pi^2=
\Biggl(
\raisebox{0.3pt}{\scriptsize $
\begin{array}{rrr} -1 & +1 & +1\\ 
+1 & -1 & +1 \\ 0  & 0 & 0
\end{array}
$}
\Biggr) ,
\label{eq:dipoleplus}
\end{equation}
with the charge $q_{g^+_{12}}\,{=}\,{+}2$ and mass  
$m_{g^+_{12}} \rum{\approx}\, 2$ and $\tilde{m}_{g^+_{12}}\rum{=}\,\infty$,
according to (\ref{eq:superpositionmass}).
If two components of the dipole are op\-po\-sit\-ely charged, say, 
$g^0_{12}\,{=}\,\Pi^1\,{+}\,\overline{\Pi}^2~$ (of whatever colour com\-bi\-nat\-ion), then
their electric fields and masses are nullified: $q_{g^0} \,{=}\, 0$,
$m_{g^0} \,{\approx}\, 0$ (but still $\tilde{m}_{g^0}\,{=}\,\infty$
due to the null-elements in the matrix $g^0$). 
The infinities in the expressions for the reciprocal masses of the dipoles
imply that neither $g^\pm$ nor $g^0$ can exist in free states
(because of their infinite energies).
However, in a large ensemble of 
neutral colour-dipoles $g^0$, not only electric but all 
the chromatic components of the field can be cancelled 
(statistically). Then, the mass of the neutral dipole $g^0_{ik}$ with 
an extra charged particle $\Pi^l$ belonging  
this ensemble but coupled to the dipole,
will be derived from the unit mass of $\Pi^l$: 
\vspace*{-5pt}
\begin{equation}
\begin{split}
m({\Pi}^i,\overline{{\Pi}}^k,{\Pi}^l)&=1\,, \\
\mbox{but} \hspace{0.2cm} \mbox{still} \hspace{0.4cm}
\tilde{m}({\Pi}^i,\overline{{\Pi}}^k,{\Pi}^l)&=\infty\,.
\label{eq:mprimgluon}
\end{split}
\end{equation}

The charge of this system will also 
be derived from the charge of the extra charged particle $\Pi^l$.

\markright{V.\,N.\,Yershov. Fermions as Topological Objects}

\section{Colour tripoles}

\markright{V.\,N.\,Yershov. Fermions as Topological Objects}

Three primitive particles with complementary colour-charges
will tend to cohere and form a {\Y}-shaped 
structure (tripole). For instance, 
by completing the set of colour-charges in the charged 
dipole [adding the blue-charged component 
to the system (\ref{eq:dipoleplus})] 
one would obtain a colour-neutral but electrically 
charged tripole: 
\vspace*{-2pt}
\begin{equation*}
{\Y}= \Pi^1+\Pi^2+\Pi^3 =
\Biggl(
\raisebox{0.3pt}{\scriptsize $
\begin{array}{rrr} -1 & +1 & +1 \\ 
+1 & -1 & +1 \\ +1 & +1 & -1   
\end{array}
$}
\Biggr) ,
\end{equation*}

\vspace*{-2pt}\noindent
which is colour-neutral at infinity but colour-polarised nearby 
(because the centres of its constituents do not coincide).
Both $m$ and $\tilde{m}$ of the tripole are finite, 
$m_{\sf Y}=\tilde{m}_{\sf Y}=3 ~[m_\circ],$
since all the diverging components of its chromofield 
are mutually cancelled (converted into the binding energy 
of the tripole).

\markright{V.\,N.\,Yershov. Fermions as Topological Objects}

\section{Doublets of tripoles}

\markright{V.\,N.\,Yershov. Fermions as Topological Objects}

\vspace*{-4mm}
\begin{figure}[htb]
\centering
\epsfig{figure=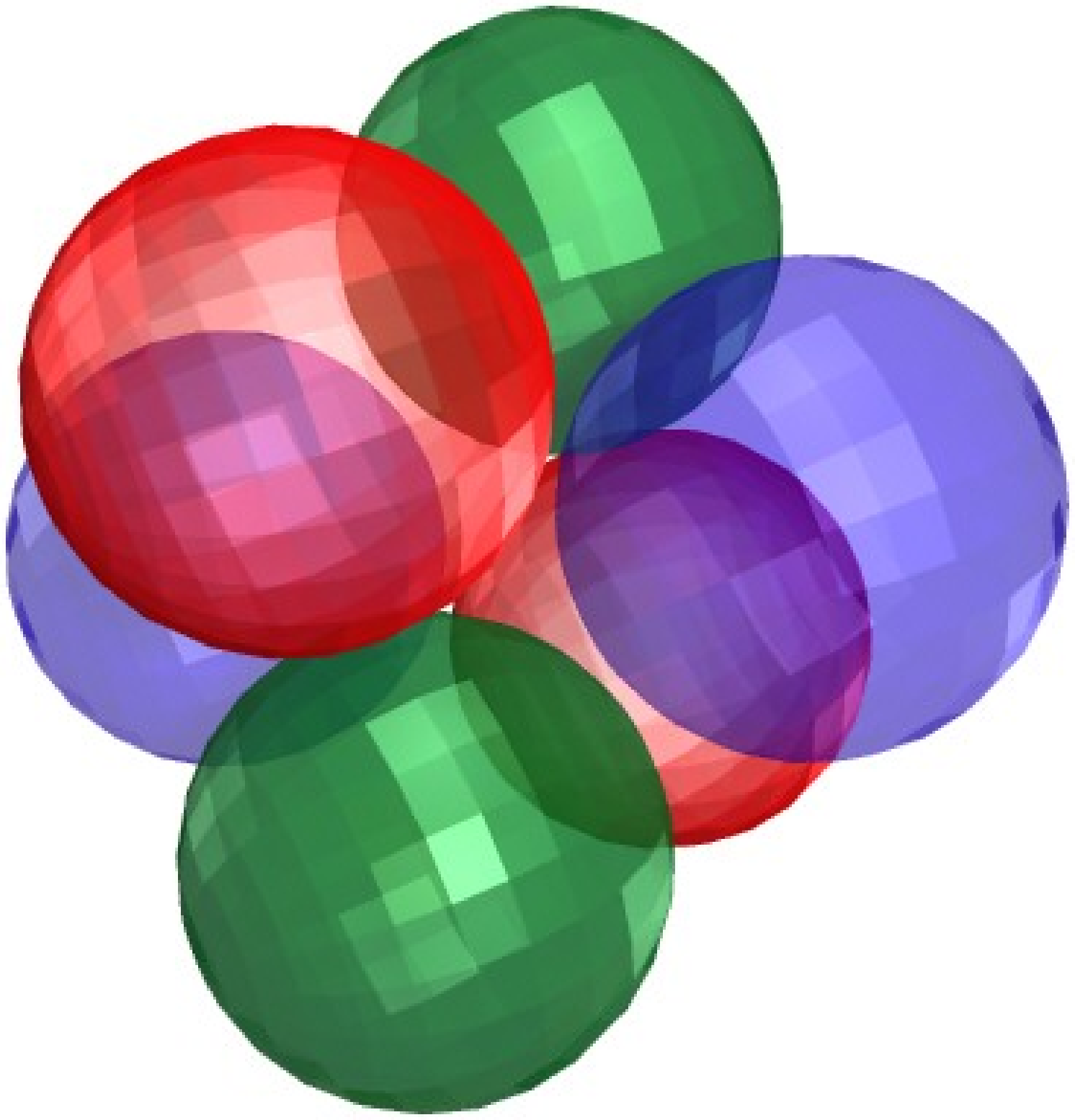,width=1.7cm} 
  \hspace{1.2cm}
\epsfig{figure=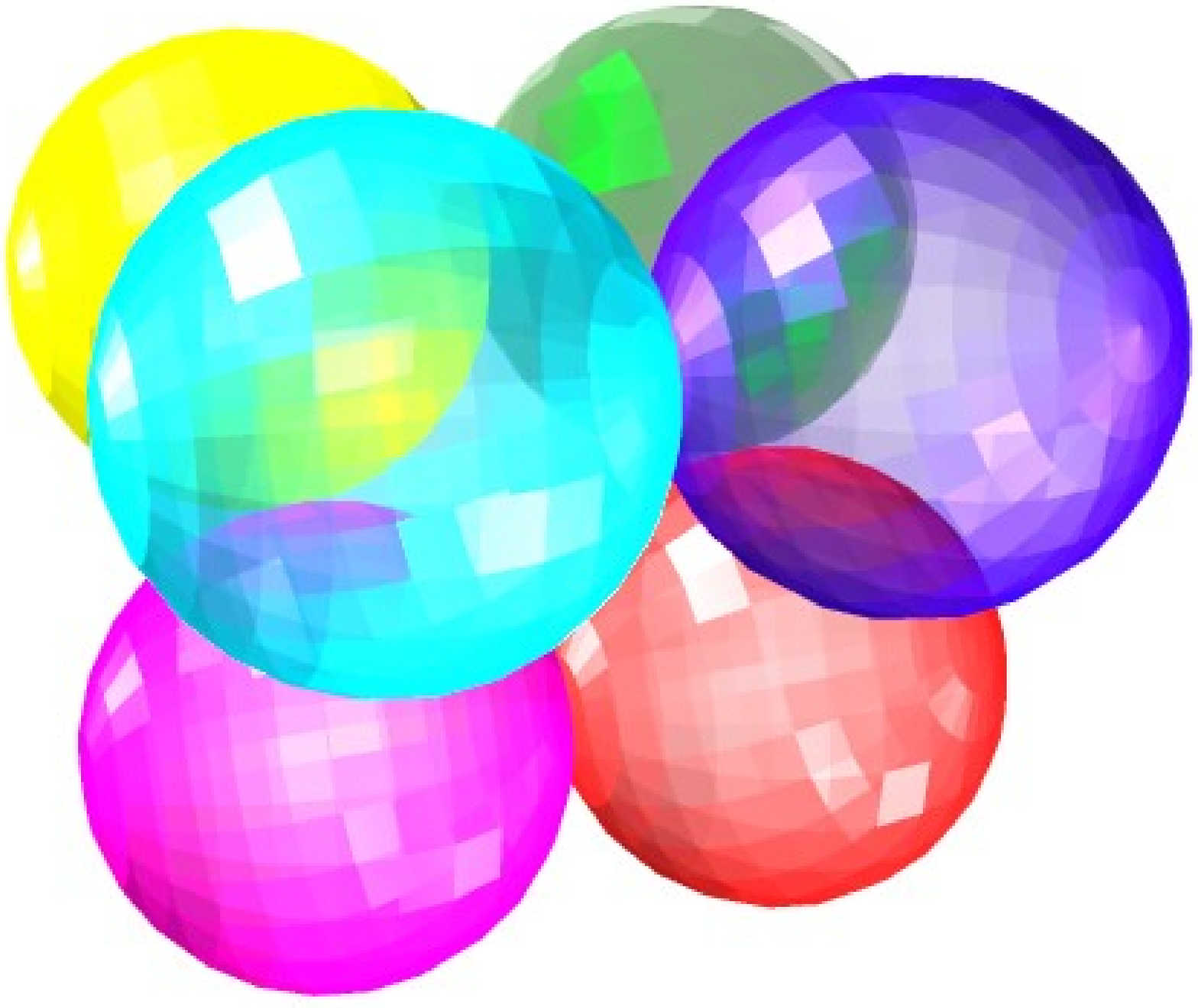,width=1.7cm} 
\put(-137,6){\scriptsize \bf {\sl a}}
\put(-52,6){\scriptsize \bf {\sl b}}
\vspace{-0.2cm}
\caption{
The tripoles ({\sf Y}-particles) can combine pairwisely, rotated 
by $180^\circ$ (a) or $120^\circ$ (b) with respect to each other.
\label{fig:twoy}
}
\vspace{-0.2cm}
\end{figure}

\vspace*{2mm}\noindent
One can show \cite{yershov05} that two like-charged {\Y}-tripoles can
combine pole-to-pole with each other and form a charged doublet 
$\delta^+=\YL \TYL$~(Fig.~\ref{fig:twoy}a). Here the rotated symbol 
\DoT{\Y} is used to indicate the rotation of the tripoles 
through $180^\circ$ with respect to each other, which corresponds to 
their equilibrium position angle. The marked arm
of the symbol ~\YL\, indicates one of the colours, say,
red, in order to visualise mutual orientations
of colour-charges in the neighbouring tripoles.  
The charge of the doublet, $q_\delta=+6~[q_\circ]$, is derived 
from the charges of its two constituent tripoles; the same is 
applied to its mass: $m_\delta=\tilde{m}_\delta=6~[m_\circ]$.  
Similarly,  if two unlike-charged {\Y}-particles are combined, 
they will form a neutral doublet, 
$\gamma=\YL \oTYL$ (Fig.\ref{fig:twoy}b)
with $q_\gamma=0$ and $m_\gamma=\tilde{m}_\gamma=0$.
The shape of the potential well in the vicinity of 
the doublet allows a certain degree of freedom
for its components to rotate oscillating
within $\pm 120^\circ$ with respect to their eq\-ui\-libr\-ium 
position angle (see \cite{yershov05} for details). 
We shall use the symbols $\circlearrowright$ and $\circlearrowleft$ to denote the
clockwise and anticlockwise rotations.

\markright{V.\,N.\,Yershov. Fermions as Topological Objects}

\section{Triplets of tripoles}

\markright{V.\,N.\,Yershov. Fermions as Topological Objects}

The $\frac{2}{3}\pi$-symmetry of the tripole allows 
up to three of them to combine if they are like-charged.
Necessarily, they will combine into a loop, denoted
hereafter with the symbol $e$. It is seen that this loop  
can be found in one of two possible configurations corresponding 
to two possible directions of rotation of the 
neighbouring tripoles: clockwise, 
$e^+_{\circlearrowright} \rub{=}\rub\, \YL\YR\YD$, and anticlockwise,  
$e^+_{\circlearrowleft} = \YL\YD\YR$.  
The vertices of the tripoles can be directed 
towards the centre of the structure (Fig.~\ref{fig:electron}a) or 
outwards (Fig.~\ref{fig:electron}b), but it is seen that 
\begin{figure}[htb]
\centering
\includegraphics[scale=0.28]{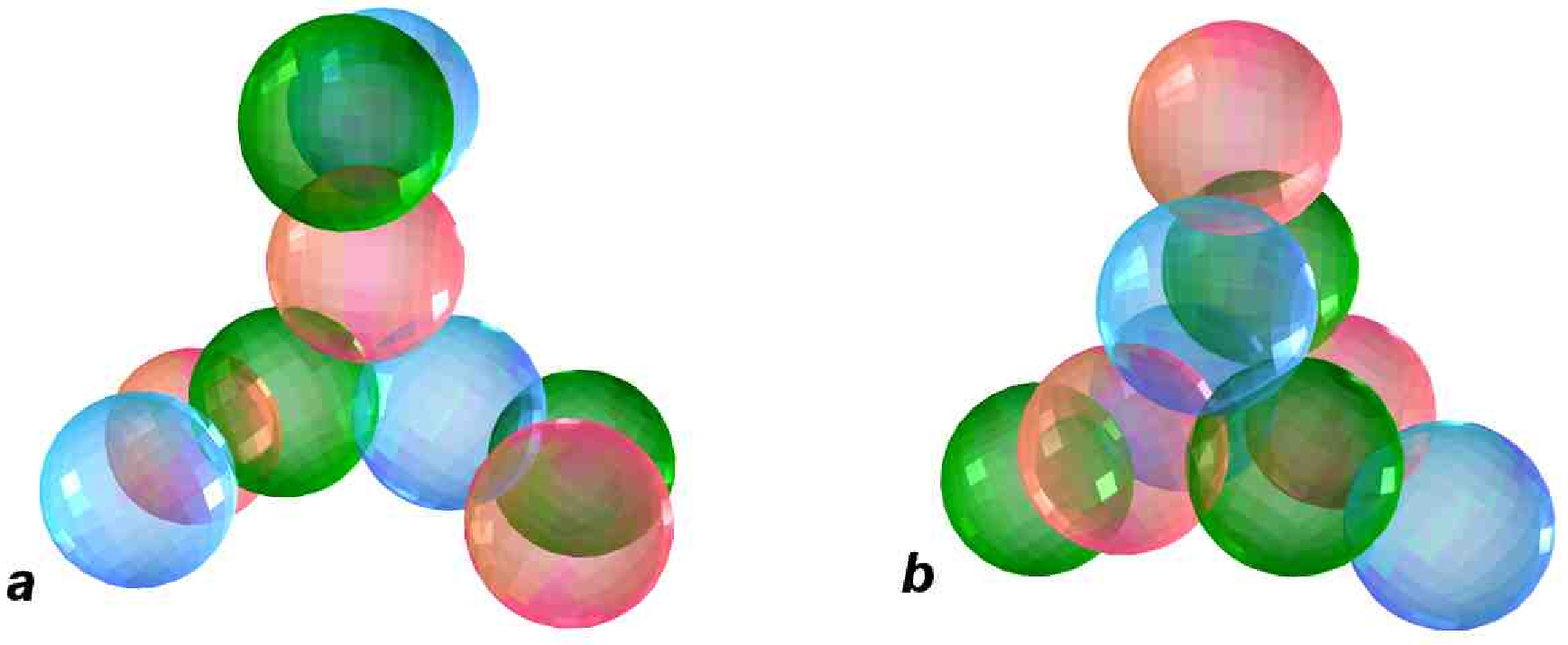}
\includegraphics[scale=0.15]{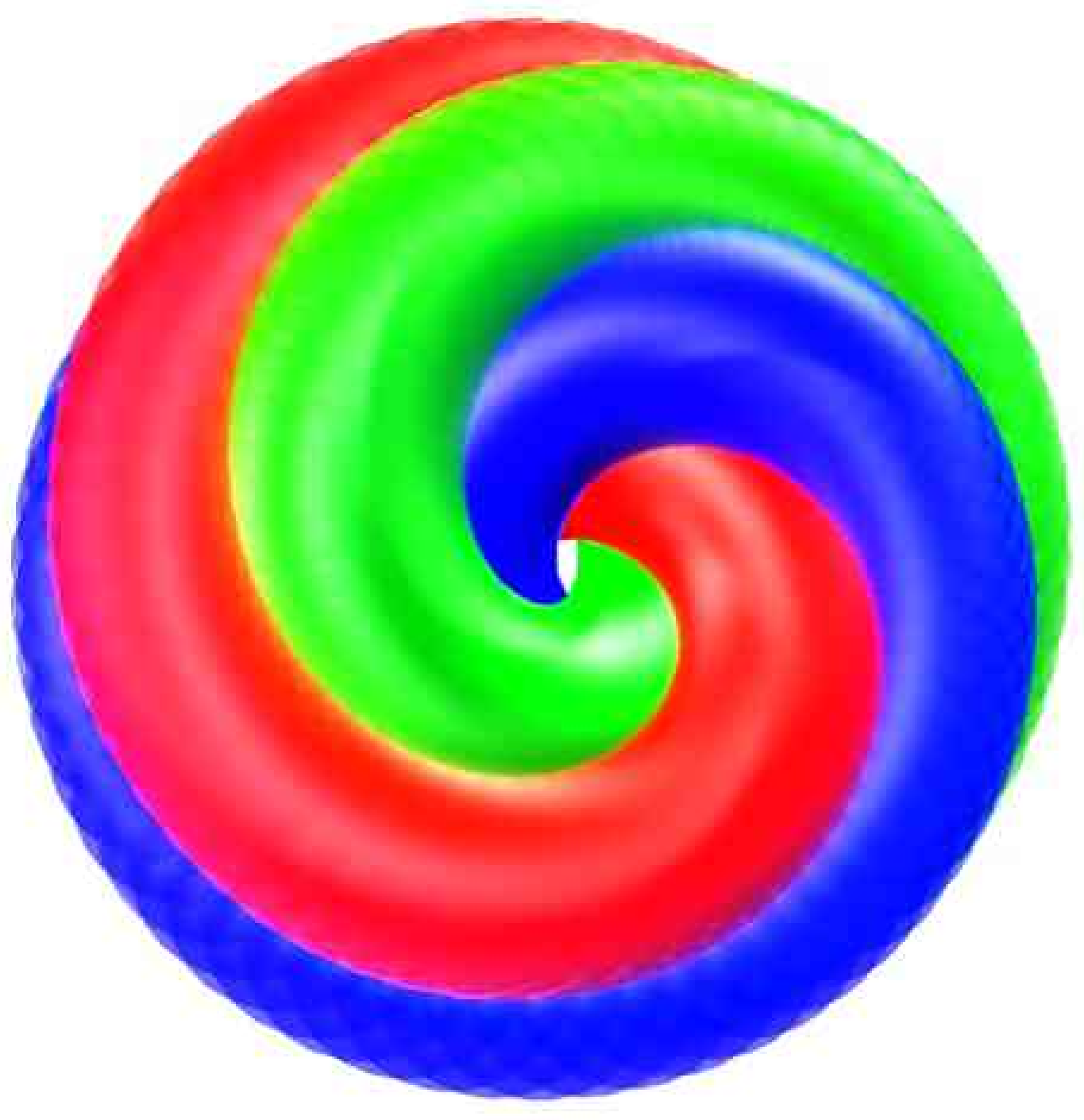}
\put(-73,4.5){\scriptsize {\bf {\sf\sl c}}}
\caption{ Three like-charged tripoles joined with their
 vertices directed towards (a) and outwards (b) of the centre 
of the structure; (c): trajectories of colour charges in this 
structure.
\label{fig:electron}
}
\end{figure}
these two orientations correspond to different 
phases of the same structure, with its colour charges  
spinning around its ring-closed axis.
These spinning charges 
will generate a toroidal (ring-closed) mag\-netic field which
will force them to move along the torus.
Their  circular motion will generate a secondary (poloidal)
magnetic field, contributing to their spin 
around the ring-axis, and so forth. 
The corresponding trajectories of colour-charges 
(currents) are shown in Fig.~\ref{fig:electron}c.
This mechanism, known as dynamo, is responsible for generating 
a self-consistent magnetic field of the triplet $e$.
 
To a first order of approximation, we shall derive
the mass of the triplet from its nine constituents,
suggesting that this mass is proportional
to the density of the currents, neglecting the contribution
to the mass of the binding and oscillatory energies of the 
tripoles. That is, we put $m_e\,{=}\,9~[m_\circ]$
(bearing in mind that the diverging components, $\tilde{m}_\circ$,
are almost null\-i\-fied). The charge of the triplet is also derived 
from the number of its constituents: $q_e\,{=}\,{\pm}\rum9~[q_\circ]$.

\markright{V.\,N.\,Yershov. Fermions as Topological Objects}

\section{Hexaplets}

\markright{V.\,N.\,Yershov. Fermions as Topological Objects}

Unlike-charged tripoles, combined pairwisely, can form\linebreak chains 
with the following patterns:
\begin{equation}
\begin{split}
\nu_{e\circlearrowright}&= 
\YL\oYR+\oTYR\TYD+\YD\,\oYL+
\oTYL\,\TYR+\YR\,\oYD+\oTYD\TYL+
\dots \\
\nu_{e\circlearrowleft}&=
\YL\oYD+\oTYD\TYR+\YR\,\oYL+
\oTYL\TYD+\YD\oYR+\oTYR\TYL+\rule[-6pt]{0pt}{0pt}
\dots
\end{split}
\label{eq:nuright}
\end{equation}
corresponding to two possible directions of rotation of the neighbouring
tripoles with respect to each other. The cycle of rotations repeats after
each  six consecutive links, making the orientation of the  
sixth link compatible with (attractive to) the first link
by the configuration of their colour-charges. This allows
the closure of the chain in a loop
(which we shall call hexaplet and denote as $\nu_e$).
 The pattern (\ref{eq:nuright}) is visualised in
Fig.~\ref{fig:twonu}a where the antipreons are coded with 
lighter colours. The corresponding trajectories of charges (currents) are
shown in Fig.~\ref{fig:twonu}b. They are clockwise or anticlockwise 
helices, similar to those of the triplet $e^-$.  The hexaplet
consists of $n_{\nu_e}\,{=}\,36$ preons (twelve tripoles); 
it is electrically neutral and, therefore, almost massless, according to
Eq.~(\ref{eq:superpositionmass}). 

Some properties of the simple preon-based structures are summarised in 
Table \ref{t:preonsummary}.
\begin{figure}[htb]
\vspace*{-2mm}
\centering
\includegraphics[scale=0.15]{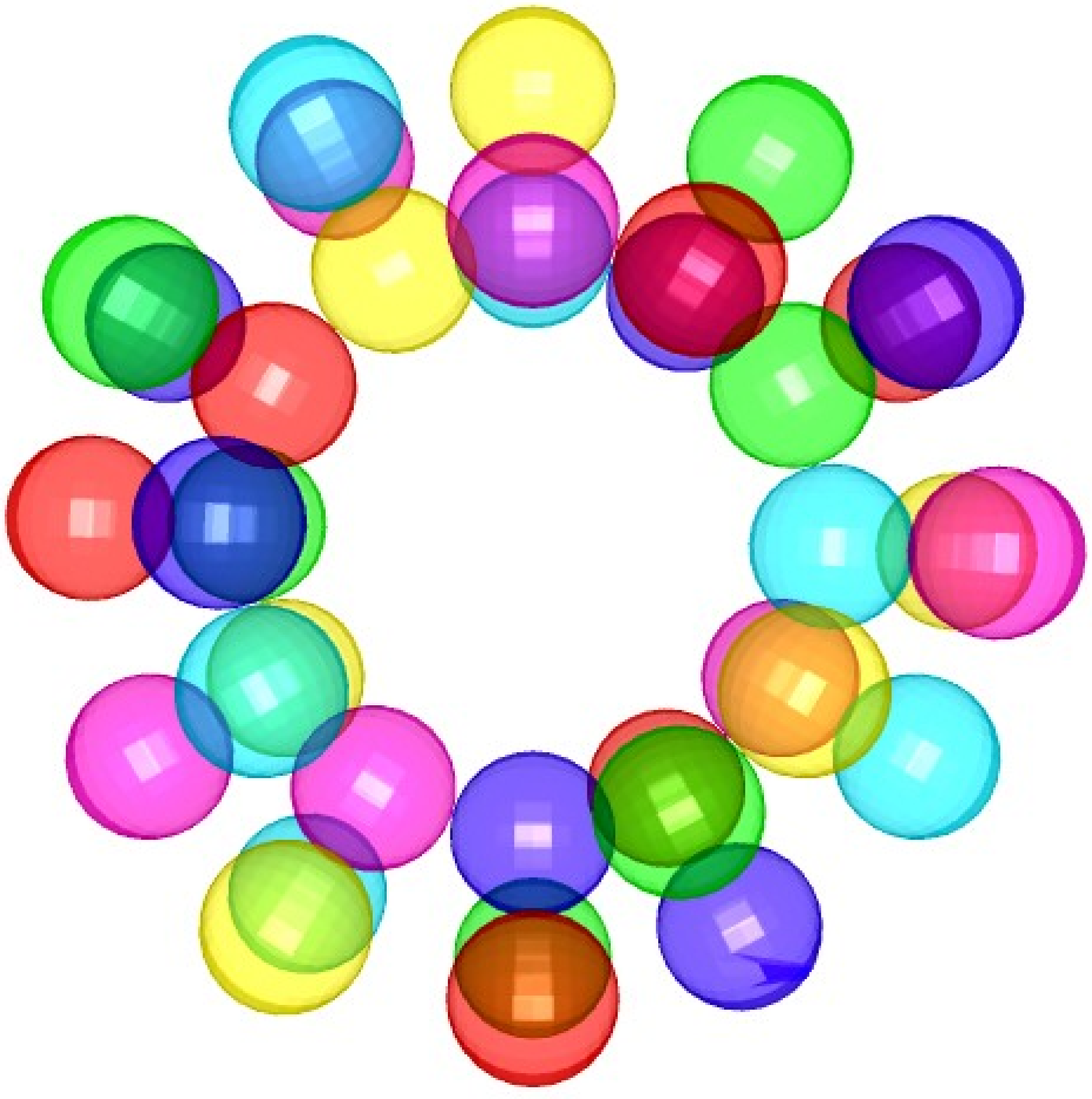}
\hspace{0.6cm}
\includegraphics[scale=0.17]{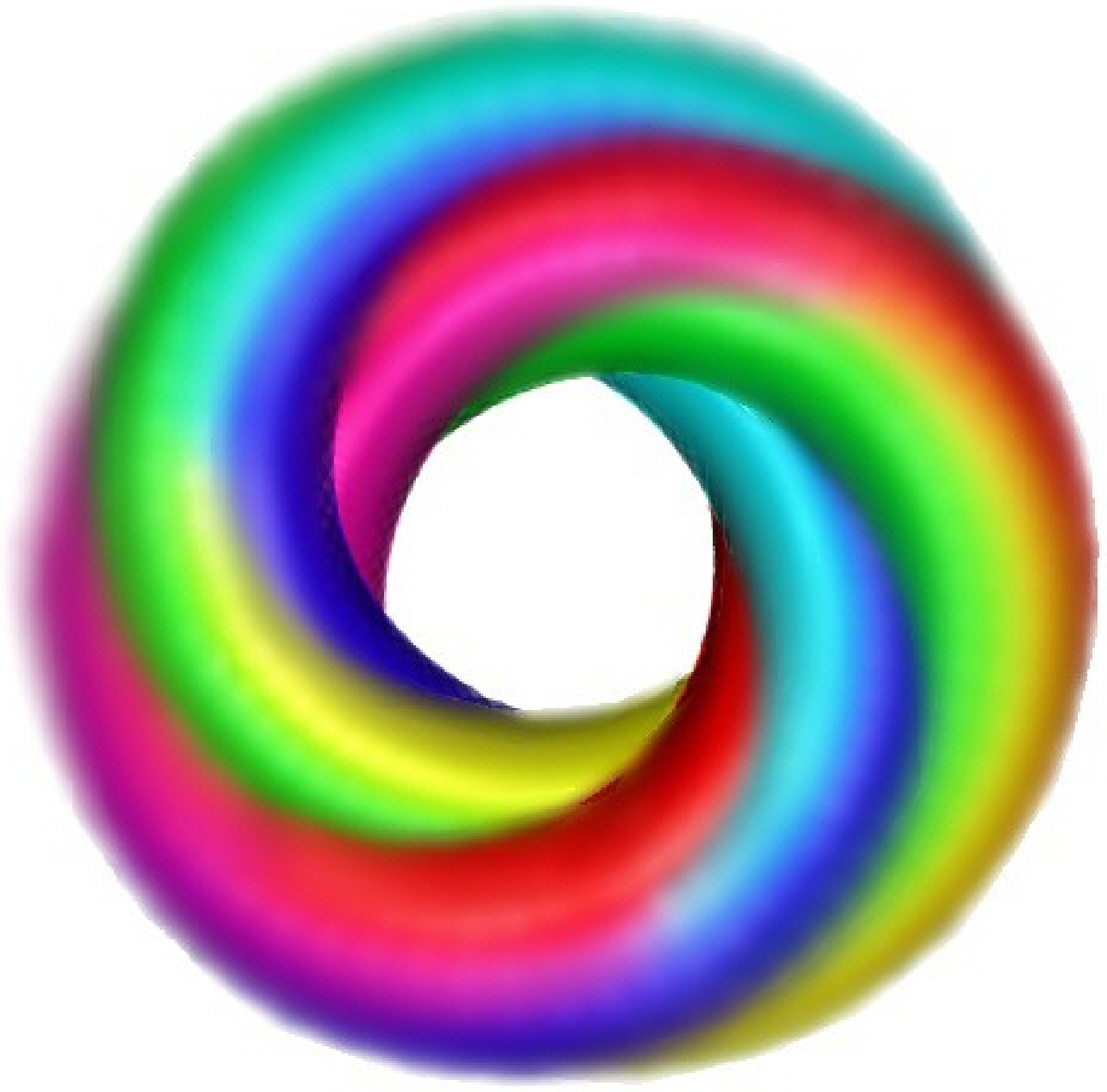}
\put(-193,7){\scriptsize {\bf {\sl a}}}
\put(-85,7){\scriptsize {\bf {\sl b}}}
\vspace{-0.2cm}
\caption{ 
{(a)}~Structure of the hexaplet $\nu_e=6{\sf Y}\overline{{\sf Y}}$ 
and {(b)}~the cor\-res\-pond\-ing helical trajectories (currents) formed by the motions 
of the hexaplet's colour-charges.
\label{fig:twonu}
}
\vspace*{-5mm}
\end{figure}

\begin{table*}[htb]
\caption{Simple preon-based structures}
\label{t:preonsummary}
\begin{center}
\small
\begin{tabular}
{|c|c|r|r|r|} \hline\hline
\parbox{1.4cm}{\small \ \ Structure}    & \parbox{2.4cm}{\small Constituents of\rule{0pt}{12pt}\\ the structure\rule[-7pt]{0pt}{0pt}} & \parbox{3.0cm}{\small Number of colour\rule{0pt}{12pt}\\ charges in the structure\rule[-7pt]{0pt}{0pt}}   & \parbox{2.6cm}{\small \ \ Charge ($q_\circ$ units)}  & 
\parbox{2.4cm}{\small \ \ Mass ($m_\circ$-units)}   \\ \hline\hline
\multicolumn{5}{|c|} {The primitive particle\rule{0pt}{11pt} 
(preon $\Pi$)\rule[-5pt]{0pt}{0pt}}                              \\ \hline
$\Pi$\rule{0pt}{10pt}              & $1\Pi$                           & 1                          &  $+1$   &   1         \\ \hline
\multicolumn{5}{|c|} {First-order structures\rule{0pt}{11pt} (combinations of preons)\rule[-5pt]{0pt}{0pt}}                              \\ \hline
$\varrho$\rule{0pt}{10pt}          & $2\Pi$                           & 2                          &  $+2$   &   2          \\ 
$g^{\mbox{\raisebox{-2.0pt}{$\scriptstyle 0$}}}$              & $1\overline{\Pi}+1\Pi$           & 2                          & $-1+1=0$  & $\thicksim 0$       \\
{\Y}            & $3\Pi$           & 3                          &  $+3$   &   3         \\  
\hline
\multicolumn{5}{|c|}{Second-order\rule{0pt}{11pt} structures (combinations of\rule[-5pt]{0pt}{0pt} tripoles {\Y})} \\
\hline
$\delta$\rule{0pt}{10pt}           & 2{\Y}                         & 6                          &  $+6$   &   6        \\ 
$\gamma$         & $1\overline{\Y}+1{\Y}$       & 6                          & $-3+3=0$ & $\thicksim 0$        \\ 
$e^-$                & 3$\overline{\Y}$                         & 9                          &  $-9$   &   9          \\ \hline
\multicolumn{5}{|c|}{Third-order\rule{0pt}{11pt} structures\rule[-5pt]{0pt}{0pt}}  \\ \hline
$2e^{-}$\rule{0pt}{10pt}   & $3\overline{\Y} + 3\overline{\Y}$             & $9+9=18$                     &  $-18$   &  18   \\
$e^{-} e^+$           & $3\overline{\Y} + 3{\Y}$ & $9+9=18$                     & $-9+9=0$  &  $\thicksim 16^\dagger$ \\
$\nu_e$            & 6$\overline{{\Y}}${\Y}     & $6\numtimes (3+3)=36$         & $6\numtimes(-3+3)=0$  & $7.9\numtimes10^{-8\,\dagger}$ \\
{\Y}$_1$        & $\nu_e \  +  \  {\Y} $ \  \   & $36+3=39$                    & $0-3=-3$ & $36+3=39$  \\
$W^-$                & $\nu_e \  +  \ e^-$  \   \       & $36+9=45$                    & $0-9=-9$ & $36+9=45$   \\
$u$                & ${{\Y}}_1$ \ \ $\between $ \ \ ${{\Y}}_1$
                                                      & $39+36+39=114$               & $+3+0+3=+6$  & $39+39=78$  \\
$\nu_\mu$          & ${\Y}_1$ \ \parbox{1mm}{\vspace*{-2.7mm}\vdots}\rule{0pt}{10.5pt}  \ \  $\overline{{\Y}}_1$ 
                                                      & $39+36+39=114$               & $-3+0+3=0$   &  $1.4 \numtimes 10^{-7\,\dagger}$  \\
$d$                & $u_{\phantom{\mu}}  +  \ W^-$                & $114+45=159$                 & $+6-9=-3$  & $78+45=123$   \\
$\mu$              & \ $\nu_{\mu}  +  \ W^-$       & $114+45=159$                 & $0-9=-9$   & $\overline{48+39}^{\,\ddagger}$ \\ 
\multicolumn{5}{|c|}{and\rule{0pt}{10pt} so\rule[-5pt]{0pt}{0pt} on\dots}  \\     
\hline\hline
\end{tabular}
\end{center}
\vspace*{-2.5mm}
{\mbox{{\footnotesize \hspace{2cm}$^\dagger$quantities estimated in \cite{yershov03}}} \par
\mbox{\raisebox{0.5mm}{{\footnotesize \hspace{2cm}$^\ddagger$system with two oscillating components (see further)}}}}
\vspace*{-3mm}
\end{table*}

\markright{V.\,N.\,Yershov. Fermions as Topological Objects}

\section{Combinations of triplets and hexaplets}

\markright{V.\,N.\,Yershov. Fermions as Topological Objects}

The looped structures $e=3{\Y}$ and  $\nu_e=6{\Y}\overline{{\Y}}$
can combine with each other, as well as with the simple tripole {\Y}, because 
of their $\frac{2}{3}\pi$-symmetry and residual chromaticism.
That is, separated from  other particles, the structure 
$\nu_e$ will behave like a neutral particle. 
But, if two such particles 
approach one another, they will be either attracted or repulsed from each other
because of van der Waals forces caused by their residual chromaticism and
polarisation. The sign 
of this interaction depends on the twisting directions of the particles'
currents. 
One can show \cite{yershov05} that the configuration of colour charges 
in the hexaplet $\nu_e$ matches (is attractive to) 
that of the triplet $e$ if both particles have like-helicities 
(topological charges). 
On the contrary, the force between the particles of the same kind
is attractive for the opposite helicities ($2e^+_{\circlearrowleft\circlearrowright}$
or $e^+_\circlearrowleft e^-_\circlearrowright$) 
and repulsive for like-helicities ($2e^+_{\circlearrowleft\circlearrowleft}$
or $e^+_\circlearrowleft e^-_\circlearrowleft$). 
So, the combined effective potential of the system $2e$ with unlike-helicities,
will have an attractive inner and repulsive outer region, allowing
an equilibrium configuration of the two particles.  
In the case of like-helicities, both inner and outer regions 
of the potential are repulsive and the particles $e$ with like-helicities
will never combine.  
This coheres with (and probably explains) the Pauli exclusion principle,
suggesting that the helicity (topological charge) of a particle 
can straight\-forwardly be related to the quantum notion of spin.
This conjecture is also supported by the fact that quantum spin is 
measured in units of angular momentum ($\hbar$), and so too ---
the topological charge in question, which is derived from the rotational
motion of the tripoles {\Y} around the ring-closed axis of the triplet $e$
or hexaplet $\nu_e$. 

Relying upon the geometrical resemblance between the 
tripoles {\Y}, triplets $e$, and hexaplets
$\nu_e$ and following the pat\-tern replicated on different 
complexity levels we can deduce 
how these structures will combine with each other. Obvious\-ly, the hexaplet
$\nu_e$, formed of twelve tripoles, is geometrically larger
than a single tripole. Thus, these two structures can com\-bine only 
\rule{-0.4pt}{0pt}when the former \rule{-0.4pt}{0pt}enfolds the \rule{-0.4pt}{0pt}latter.
 The combin\-ed struc\-ture, which we shall denote as
%
${\Y}_1=\nu_e+{\Y},$
%
 will have a mass derived from its 39 constituents: 
$m_{{\Y}_1}^{\phantom{0}}\,{=}\, n_{\nu_e}\,{+}\rub\,m_{\Y}^{\phantom{0}}
\,{=}$\linebreak ${=}\,36\,{+}\,3\,{=}\,39~[m_\circ]$. 
Its charge will be derived from the charge of its central 
tripole: $q_{{\,\Y}_1}^{\phantom{0}}\,{=}\,{\pm}\rub 3~[q_\circ]$. By their prop\-ert\-ies,
the tripole, {\Y}, and the ``helical tripole'', ${\Y}_1$, 
are alike, except for the helicity property of 
the latter derived from the helicity of its constituent hexaplet. 

When considering the combination of the hexaplet, $\nu_e$, with the triplet, $e$,
we can observe that the hexaplet must be stiffer than the triplet because of 
stronger bonds between the unlike-charged components of the former, while
the repulsion between the like-charged components of the latter makes the
bonds between them weaker. Then, the amplitude of the 
fluctuations of the triplet's radius will be larger than 
that of the hexaplet. Thus, in the combined structure, which we shall denote as
$W\,{=}\,6{\Y}\overline{\Y}3{\Y}$ (or $\nu_ee$), it is the triplet that would 
enfold the hexaplet. 
The charge of this structure will \hspace{-0.2ex}correspond to the charge of its charged 
component, $e$: $q_W^{\phantom{0}}\,{=}$
${=}\,{\pm}\rub9~[q_\circ]$; its mass can also be derived 
from the masses of its constituents if oscillations are dampened: 
$$
m_{W}=m_e+ n_{\nu_e}=9+36=45~[m_\circ].
$$

Like the simple {\Y}-tripoles, the ``helical''
ones, ${\Y}_1$, can form bound states with each other (doublets,
strings, loops, etc.).
Two hexaplets, if both enfold like-charged tripoles, will {\it always} 
have like-topological charges (helicities), which means that the force 
between them due to their topological charges will be repulsive 
(in addition to the usual repulsive force between like-charges).
Thus, two like-charged helical tripoles ${\Y}_1$
will never combine, unless there exists an inter\-mediate 
hexaplet ($\nu_e$) between them, with the topological charge 
opposite to that of the components of the pair.
This would neutralise the repulsive force between these com\-ponents and
allow the formation of the following positively charged bound state 
(``helical'' doublet):
\begin{equation}
u^+={\Y}_{1\circlearrowright} \nu_{e\circlearrowleft} {\Y}_{1\circlearrowright} 
\hspace{0.4cm} {\rm or} 
\hspace{0.4cm} {\Y}_1 \between {\Y}_1~.
\label{eq:uplink2a}
\end{equation}

For brevity we have denoted the intermediate hexaplet with 
the symbol $\between$, implying that it creates a bond force between 
the otherwise repulsive components on its sides.
By its properties, the helical doublet can be identified with the {\it u}-quark.
Its net charge, $q_u=+6~[q_\circ]$, is derived from the charges of its two charged 
components (${\Y}_1$-tripoles). Its mass is also derived from  
the number of particles that constitute these charged components:
$m_u\,{=}\,2\numtimes 39\,{=}\,78~[m_\circ].$ 
The positively charged $u$-quark can combine with the neg\-at\-ively charged
structure $W^-\,{=}\:\overline{\nu}_ee^-$ (of 45-units mass),
forming the {\it d}-quark:
\begin{equation}
d^-=u^++\,\overline{\nu}_ee^-
\label{eq:dlink2a}
\end{equation}
of a 123-units mass ($m_d=m_u+m_{W}=78+45)$. The charge of this structure
will correspond to the charge of a single triplet: 
$q_d=q_u+q_e=+6-9=-3~[q_\circ]$ (see Fig.~\ref{fig:down}).
\newcommand{\Neutrino}{%
 \put(0,10){\arc(2,0){180}}
 \put(0,-10){\arc(2,0){-180}}
 \put(-2,-10){\line(0,1){20}}
 \put(2,-10){\line(0,1){20}} }
\newcommand{\RTriune}{%
 \put(2.2,0.0){\curve(0.0,7.0, 0.5,3.0, 2.0,0.0)}
 \put(2.2,0.0){\curve(0.0,-7.0, 0.5,-3.0, 2.0,0.0)}
 \put(2.2,-7.0){\line(0,1){14} } }
\newcommand{\LTriune}{%
 \put(-2.2,0.0){\curve(0.0,7.0, -0.5,3.0, -2.0,0.0)}
 \put(-2.2,0.0){\curve(0.0,-7.0, -0.5,-3.0, -2.0,0.0)}
 \put(-2.2,-7.0){\line(0,1){14} } }
\newcommand{\Electron}{%
 \put(0.0,0.0){\curve(0.0,10.0, -1.0,4.5, -2.0,0.0)}
 \put(0.0,0.0){\curve(0.0,-10.0, -1.0,-4.5, -2.0,0.0)}
 \put(0.0,0.0){\curve(0.0,10.0, 1.0,4.5, 2.0,0.0)}
 \put(0.0,0.0){\curve(0.0,-10.0, 1.0,-4.5, 2.0,0.0)} }
%
\begin{figure}[htb]
\hspace{1.5cm}
\setlength{\unitlength}{0.6mm}
\begin{picture}(100,50)(-10,10)
 \put(-2,50.5){\makebox(0,0)[t]{\scriptsize Charge:}}
 \put(18,58){\makebox(17,0)[t]{\tiny $\stackrel{{\rm Net
     \hspace{0.1cm} charge}  \hspace{0.1cm}  -3} {\overbrace{ 
       \mathbf{-9} \hspace{0.12cm}
     \mathbf{+3} \hspace{0.8cm} \mathbf{+3}}}$}}
 \put(-9,23){\makebox(0,0)[t]{\scriptsize Number}}
 \put(-7,20){\makebox(0,0)[t]{\scriptsize of charges:}}
 \put(16,20){\makebox(18,0)[t]{\tiny $\stackrel{\underbrace{
    \mathbf{36} \hspace{0.1cm} \mathbf{9}}}{\nu_e \hspace{0.1cm} e^-}
    \hspace{0.08cm} \stackrel{\underbrace{\mathbf{3} \hspace{0.03cm}
    \mathbf{36} \hspace{0.06cm} (36) \hspace{0.1cm} \mathbf{36}
    \hspace{0.06cm} \mathbf{3}}}{u^+}$ }}
 \put(10,34){\Neutrino}
 \put(14,34){\Electron}
 \put(22,34){\Neutrino}
 \put(30,34){\Neutrino}
 \put(38,34){\Neutrino}
 \put(38,34){\RTriune}
 \put(22,34){\LTriune}
 \put(8.5,33){\scriptsize $\circlearrowleft$}
 \put(12.5,33){\scriptsize $\circlearrowleft$}
 \put(20.5,33){\scriptsize $\circlearrowright$}
 \put(28.5,33){\scriptsize $\circlearrowleft$}
 \put(36.5,33){\scriptsize $\circlearrowright$}
 \put(11,7){\makebox(0,0)[t]{\scriptsize (mass 45)}}
 \put(32,7){\makebox(0,0)[t]{\scriptsize (mass 78)}}
\end{picture}
\vspace{0.1cm}
\caption{Scheme of the $d$-quark. The symbol
$\diamondsuit$ is used for the triplet ($e$), the symbols $\langle 
\hspace{-0.8mm} \mid$ and $\mid \hspace{-0.8mm} \rangle$ 
denote the tripoles ({\sf Y}-particles), and the symbols
$\genfrac{}{}{0pt}{3}{\cap}{\cup}$\hspace{-1.5ex}$\shortmid\hspace{0.3ex}\shortmid$
denote the hexaplets ($\nu_e$).
 \label{fig:down}
}
\vspace*{-3mm}
\end{figure}
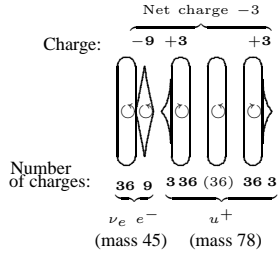

\markright{V.\,N.\,Yershov. Fermions as Topological Objects}

\vspace*{-3mm}
\section{The second and third generations of the fundamental\protect\linebreak
\protect\rule{1.2pt}{0pt}fermions}

\markright{V.\,N.\,Yershov. Fermions as Topological Objects}

When two unlike-charged helical tripoles combine, 
their  po\-lar\-isation modes and helicity signs will {\it always} be 
opposite (simply because their central tripoles have  
opposite charges). This would cause an attractive force between 
these two part\-icles, in addition to the usual attractive force corresponding to 
the opposite electric charges of ${\Y}_1$ and $\overline{\Y}_1$. 
Since all the forces here are attractive, the components of this system 
will coalesce and then disintegrate into neutral doublets $\gamma$.
However, this coalescence can be prevented by an additional hexaplet $\nu_e$
with oscillating polarisation, which would create a repulsive stabilising force 
(barrier) between the combining particles:
%
%
\begin{equation}
\nu_\mu= 
{\Y}_{1\circlearrowright} \nu_{e\circlearrowright\circlearrowleft} 
\overline{\Y}_{1\circlearrowleft} . 
 \label{eq:numu}
\end{equation}

It is natural to identify this structure with the muon-neutrino --- 
a neutral lepton belonging to the second family of the fundamental fermions.
The intermediate hexa\-plet os\-cil\-lat\-es between the tripoles 
${\Y}_{1\circlearrowright}$ and $\overline{\Y}_{1\circlearrowleft}$,
changing synchron\-ously its polarisation state:
%
$\nu_{e\circlearrowright} \leftrightsquigarrow \nu_{e\circlearrowleft}\,.$
%
%
For brevity, we shall use vertical dots separating the
components of $\nu_\mu$ to denote this barrier-hexaplet:
\vspace{-0.2cm}
\begin{equation}
\nu_\mu = {\Y}_1 \vdots \hspace{0.5ex} \overline{\Y}_1\,.
\label{eq:nu_mu0}
\end{equation}

By analogy, we can derive the tau-neutrino structure:
%
\begin{equation}
\nu_\tau = {\Y}_1 \vdots \hspace{0.5ex} \overline{\Y}_1
\vdots \hspace{0.5ex} {\Y}_1 \vdots \hspace{0.5ex} \overline{\Y}_1\,,
\label{eq:nu_tau0}
\end{equation}
as well as the structures of the muon (Fig.~\ref{fig:muon}): 
%
\begin{equation}
\mu^-=\nu_\mu \overline{\nu}_e e^- 
\label{eq:muon}
\end{equation}
and tau-lepton (Fig.~\ref{fig:tau}):
%
\begin{equation}
\tau^-=\nu_\tau \overline{\nu}_\mu \mu^- .
\label{eq:tau0}
\end{equation}

Drawing also an analogy with molecular equilibrium configurations, 
where the rigidness of a system depends on the number of local minima 
of its combined effective potential \cite{burenin02}, we can consider 
the second and third gen\-er\-at\-ion fermions as non-rigid structures 
with oscillating com\-po\-nents (clusters) rather than stiff entities 
with dampened oscillations. In Fig.~\ref{fig:muon} and Fig.~\ref{fig:tau}
we mark the supposedly clustered components 
of the $\mu$- and $\tau$-leptons with braces. 
Obtaining the ground-state energies (masses) of these 
com\-plex structures is not a straightforward task because 
they may have a great variety of 
oscillatory modes contributing to the mass.
However, in principle, these masses are computable, as can be shown
by using the following empirical formula: 
\begin{equation}
m_{\rm clust}=
\overline{m_1+m_2+\dots +m_N}=m\tilde{m},
\label{eq:mtotal}
\end{equation}
where $N$ is the number of oscillating clusters, each with the   
mass $m_i$ ($i=1, \dots, N$); $m$ is the sum of these masses:
\begin{equation*}
m=m_1+m_2+\dots +m_N,
\end{equation*}
and\, $\tilde{m}$ is the reduced mass based on the components (\ref{eq:superpositionmass}):
\begin{equation*}
\tilde{m}^{-1}={\tilde{m}_1^{-1}+\tilde{m}_2^{-1}+\dots +\tilde{m}_N^{-1}}.
\end{equation*}

For simplicity, we assume that unit conversion 
coef\-fi\-ci\-ents in this formula are set to unity.
%
Each substructure here contains a well-defined number of constituents
(preons) corresponding to the configuration with the lowest energy.
Therefore, the number of these constituents 
is fixed by the basic symmetry of the potential, implying
that the input quantities in (\ref{eq:mtotal}) are not free parameters. 
The fermion masses computed with the use of this formula
are summarised in Table~\ref{t:massresult}.

\begin{figure}[htb]
\hspace{1.5cm}
\setlength{\unitlength}{0.6mm}
\begin{picture}(100,50)(-10,10)
 \put(-2,52){\makebox(0,0)[t]{\scriptsize Charge:}}
 \put(15,58.5){\makebox(0,0)[t]{\tiny $\stackrel{q_1=-6}
 {\overbrace{ \mathbf{-9} \hspace{0.3cm}
     \mathbf{+3}}}$}}
 \put(37,58.5){\makebox(0,0)[t]{\tiny $\stackrel{q_2= -3}
  {\overbrace{\mathbf{-3}}}$}}
 \put(-6,24.2){\makebox(0,0)[t]{\scriptsize Number}}
 \put(-3.2,20.2){\makebox(0,0)[t]{\scriptsize of charges:}}
 \put(15,21){\makebox(0,0)[t]{\tiny $\stackrel{\underbrace{
    \mathbf{45} \hspace{0.2cm}  \mathbf{3}}}
    {e^- \hspace{0.08cm} \nu_e \hspace{0.15cm} {\Y}}$}}
 \put(27,20.5){\makebox(0,0)[t]{\tiny 36 \hspace{0.05cm} 36}}
 \put(38,21){\makebox(0,0)[t]{\tiny $\stackrel{\underbrace{
    \mathbf{3} \hspace{0.1cm} \mathbf{36}}}{ \ \ \overline{\Y}_1}$}}
 \put(15,35){\Neutrino}
 \put(11,35){\Electron}
 \put(22,35){\Neutrino}
 \put(30,35){\Neutrino}
 \put(38,35){\Neutrino}
 \put(38,35){\LTriune}
 \put(21,35){\LTriune}
 \put(9.5,34){\scriptsize $\circlearrowright$}
 \put(13.5,34){\scriptsize $\circlearrowright$}
 \put(20.5,34){\scriptsize $\circlearrowright$}
 \put(29.4,34){\scriptsize $\wr$}
 \put(36.5,34){\scriptsize $\circlearrowleft$}
 \put(14.5,8.5){\makebox(0,0)[t]{\scriptsize ($m_1=48$)}}
 \put(37,8.5){\makebox(0,0)[t]{\scriptsize ($m_2=39$)}}
\end{picture}
\caption{Scheme of the muon. 
 \label{fig:muon}
}
\vspace{0.1cm}
\end{figure}
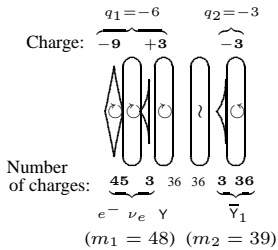

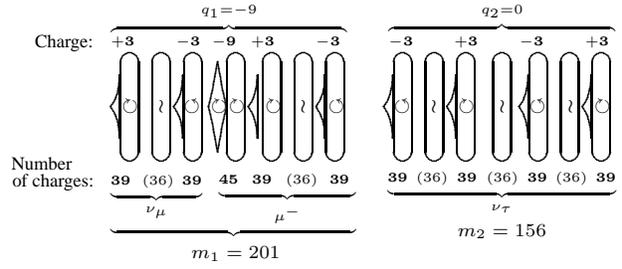
\begin{figure}[htb]
\centering
\setlength{\unitlength}{0.6mm}
\begin{picture}(100,50)(-15,10)
 \put(-25,51){\makebox(0,0)[t]{\scriptsize Charge:}}
 \put(11.5,58){\makebox(0,0)[t]{\tiny $\stackrel{q_1=-9}
 {\overbrace{\mathbf{+3} \hspace{0.55cm} \mathbf{-3} \hspace{0.15cm} \mathbf{-9} \hspace{0.2cm}
     \mathbf{+3} \hspace{0.55cm} \mathbf{-3}~~}}$}}
 \put(72,58){\makebox(0,0)[t]{\tiny $\stackrel{q_2=0}
  {\overbrace{\mathbf{-3} \hspace{0.55cm} \mathbf{+3} \hspace{0.55cm} \mathbf{-3} 
  \hspace{0.55cm} \mathbf{+3}~~}}$}}
 \put(-30,24){\makebox(0,0)[t]{\scriptsize Number}}
 \put(-27.5,20.2){\makebox(0,0)[t]{\scriptsize of charges:}}
 \put(12.5,21){\makebox(0,0)[t]{\tiny $\stackrel{
 \underbrace{
             \stackrel{\underbrace{
                                   \mathbf{39} \hspace{0.15cm} (36) \hspace{0.1cm} \mathbf{39}
                                  }
                      }{\nu_\mu} \hspace{0.15cm}
             {\raisebox{-0.1cm}{$\stackrel{\underbrace{
                                   \mathbf{45} \hspace{0.18cm} \mathbf{39} 
                                   \hspace{0.18cm} (36) \hspace{0.15cm} \mathbf{39~~}
                                  }
                       }{\mu^-}$}}
             }                                      }{ \hspace{0.1cm}}$}}
 \put(72,21){\makebox(0,0)[t]{\tiny $\stackrel{\underbrace{
    \mathbf{39} \hspace{0.11cm} (36) \hspace{0.11cm} \mathbf{39} \hspace{0.13cm} 
    (36) \hspace{0.11cm} \mathbf{39} \hspace{0.11cm} (36) \hspace{0.11cm} 
    \mathbf{39}}}{\nu_\tau}$}}
 \put(13,35){\Neutrino}
 \put(9,35){\Electron}
 \put(21,35){\Neutrino}
 \put(28,35){\Neutrino}
 \put(35,35){\Neutrino}
 \put(35,35){\LTriune}
 \put(20,35){\LTriune}
 \put(7.5,34){\scriptsize $\circlearrowright$}
 \put(11.5,34){\scriptsize $\circlearrowright$}
 \put(19.5,34){\scriptsize $\circlearrowright$}
 \put(27.3,34){\scriptsize $\wr$}
 \put(33.5,34){\scriptsize $\circlearrowleft$}
 \put(50,35){\Neutrino}
 \put(57,35){\Neutrino}
 \put(64,35){\Neutrino}
 \put(72,35){\Neutrino}
\put(80,35){\Neutrino}
 \put(87,35){\Neutrino}
 \put(94,35){\Neutrino}
 \put(50,35){\LTriune}
 \put(64,35){\LTriune}
 \put(80,35){\LTriune}
 \put(94,35){\LTriune}
 \put(48.0,34){\scriptsize $\circlearrowleft$}
 \put(56.2,34){\scriptsize $\wr$}
 \put(62.3,34){\scriptsize $\circlearrowright$}
 \put(71.4,34){\scriptsize $\wr$}
 \put(78.0,34){\scriptsize $\circlearrowleft$}
 \put(86.5,34){\scriptsize $\wr$}
 \put(92.5,34){\scriptsize $\circlearrowright$}
 \put(13,4){\makebox(0,0)[t]{\scriptsize $m_1=201$}}
 \put(72,9){\makebox(0,0)[t]{\scriptsize $m_2=156$}}
 \put(-10.5,35){\Neutrino}
 \put(-3.5,35){\Neutrino}
 \put(3.5,35){\Neutrino}
 \put(3.5,35){\LTriune}
 \put(-10.5,35){\LTriune}
 \put(-12.2,34){\scriptsize $\circlearrowright$}
\put(-4.2,34){\scriptsize $\wr$}
 \put(1.5,34){\scriptsize $\circlearrowleft$}
%
\end{picture}
\vspace*{3mm}
\caption{Scheme of the tau-lepton. 
 \label{fig:tau}
}
\vspace{-0.1cm}
\end{figure}

As an  example, let us compute the muon's mass. The mass\-es of the muon's
substructures, according to Fig.~\ref{fig:muon},~are: 
$m_1\,{=}\,\tilde{m}_1\,{=}\,48$, $m_2\,{=}\,\tilde{m}_2\,{=}\,39$ (in units of $m_\circ$). 
And the muon's mass will be:
$m_\mu \,{=}\, \overline{48{+}39}\rub{=}\rub
\frac{48+39}{1/48+1/39}\rub{=}\rub 1872 ~[m_\circ]$.
For the $\tau$-lepton, the constituent masses are $m_1=\tilde{m}_1\,{=}\,201$,
$m_2\,{=}\,\tilde{m}_2\,{=}\,156$ (Fig.~\ref{fig:tau}), and its mass is 
$m_\tau  \,{=}\, \overline{201{+}156}\,{=}$\linebreak
${=}\,31356 ~[m_\circ].$
For the proton, the positively charged fermion consisting of two up 
($N_u\,{=}\,2$), one down ($N_d\,{=}\,1$) quarks and submerged into 
a cloud of gluons $g^0$, the masses
of its components 
are $m_u\,{=}\,\tilde{m}_u\,{=}\,78$,  $m_d\,{=}\,\tilde{m}_d\,{=}\,123$.
The total number of primitive charges comprising the 
proton's struc\-ture is 
%
$N_p\,{=}\,2\rub m_u +m_d\,{=}\,2 \numtimes 78+123
\,{=}\,279,$
%
which would correspond to the number of gluons ($N_g$) interacting with 
each of these charges ($N_g\,{=}\,N_p\,{=}\,279$).
The masses of these gluons, according 
to (\ref{eq:mprimgluon}), are
$m_{g^0}\,{=}\,1$, 
$\tilde{m}_{g^0}\,{=}\,\infty$, and the resulting proton mass is 
\begin{equation}
m_p  = \overline{N_u\rub m_u+N_d\,m_d+N_g\rub m_g}= 16523 ~[m_\circ]\,,
\label{eq:pmass}
\end{equation}
which also reproduces the well-known but not yet explained 
proton-to-electron mass ratio, since 
$\frac{m_p}{m_e}\,{=}\,\frac{16523}{9} \,{\approx }\,1836$.
%
 
\begin{table*}[htb]
\caption{Computed masses of quarks and leptons. 
The values in the 4th column taken in units of $m_\circ$ 
are converted into proton mass units (5th column) 
$m_p$=16523, Eq.(\ref{eq:pmass}). The overlined ones are shorthands for Eq.~(\ref{eq:mtotal}).
The masses of $\nu_e$, $\nu_\mu$ and $\nu_\tau$ are estimated in \cite{yershov03}.}
\label{t:massresult}
\begin{center}
\small
\begin{tabular}
{|c|c|l|l|l|l|} \hline\hline
\multicolumn{2}{|c|}{\parbox{3.0cm}{Particle and its\rule{0pt}{12pt}\\ structure (components)\rule[-6pt]{0pt}{0pt}}}  & \parbox{3.9cm}{Number of charges\rule{0pt}{12pt} in the non-cancelled mass components\rule[-6pt]{0pt}{0pt}}  &  \parbox{2.4cm}{Computed  masses\rule{0pt}{12pt}  in units of $[m_p]$\rule[-6pt]{0pt}{0pt}}  & \parbox{2.4cm}{Masses converted\rule{0pt}{12pt} into $m_p$\rule[-6pt]{0pt}{0pt}}    & \parbox{3.2cm}{Experimental masses\rule{0pt}{12pt} \cite{properties}
in units of $[m_p]$\rule[-6pt]{0pt}{0pt}} \\ \hline\hline
\multicolumn{6}{|c|} {First\rule{0pt}{11pt} family\rule[-5pt]{0pt}{0pt}} \\ \hline
$\nu_e$ & $6{\Y}\overline{{\Y}}$\rule{0pt}{10pt}            & $\approx 0$   & $7.864\numtimes 10^{-8}$ & $4.759\numtimes 10^{-12}$  & $<3 \numtimes 10^{-9}$ \\
$e^-$ & $3\overline{{\Y}}$               & 9    & 9    & 0.0005447  & 0.0005446170232 \\ 
$u$ & $ {\Y}_1 \hspace{0.2cm} \between \hspace{0.2cm} {\Y}_1$ & 78  & 78 & 0.004720 & 0.0021 to 0.0058 \\
$d$ & $u$ $ \hspace{0.2cm} + \hspace{0.2cm} \overline{\nu}_e e^-$                 & 123  & 123  & 0.007443 & 0.0058 to 0.0115 \\
\hline
\multicolumn{6}{|c|}{Second\rule{0pt}{11pt} family\rule[-5pt]{0pt}{0pt}} \\
\hline
$\nu_\mu $ & ${\Y}_1$ \ \parbox{1mm}{\vspace*{-2.6mm}\vdots}\rule{0pt}{10.5pt}  \ \  $\overline{{\Y}}_1$ & $\approx 0$  & $1.4\numtimes 10^{-7}$ & $8.5\numtimes 10^{-12}$ 
&  $<2 \numtimes 10^{-4}$\\
$\mu^-$ & \ \ $\nu_\mu$ \ \ + \   $\overline{\nu}_e e^-$    & $\overline{48+39}$   & 1872 & 0.1133 & 0.1126095173 \\
$c$ & ${{\Y}_2}$ \ \ $\between$ \ \ ${{\Y}_2}$ & $\overline{165+165}$ & 27225  & 1.6477 & 1.57 to 1.95 \\
$s$ &  \ $c$ \ \ \ + \ \ $e^-$ & $\overline{165+165+9}$ & 2751 & 0.1665 & 0.11 to 0.19 \\
\hline
\multicolumn{6}{|c|}{Third\rule{0pt}{11pt} family\rule[-5pt]{0pt}{0pt}} \\
\hline
$\nu_\tau$ & ${\Y}_1$\,\parbox{1mm}{\vspace*{-2.6mm}\vdots}\rule{0pt}{10.5pt} $\overline{{\Y}}_1$\,\parbox{1mm}{\vspace*{-2.6mm}\vdots}%
\,${\Y}_1$\parbox{1mm}{\vspace*{-2.6mm}\vdots} $\overline{{\Y}}_1$
 & $\approx 0$  
& $1.5896\numtimes10^{-7}$ & $9.6192\numtimes10^{-12}$ & $<2 \numtimes 10^{-2}$ \\
$\tau^-$ & \ \ $\nu_\tau$ \ \  + \    $\overline{\nu}_\mu \mu^-$ & $\overline{156+201}$ & 31356 & 1.8977 & $1.8939\pm 0.0003$ \\
$t$ & ${{\Y}_3}$ \  $\between$ \ \ ${{\Y}_3}$ & $\overline{1767+1767}$ & 3122289 & 
188.94 & $189.7\pm 4.5$ \\
$b$ & \  $t$ \ \ + \ \ $\mu^-$ & $\overline{1767+1767+48+39}$ & 76061.5 & 4.603 & 4.3 to 4.7 \\
\hline\hline
\end{tabular}
\end{center}
\end{table*}

With the value (\ref{eq:pmass}) one can convert $m_e$, $m_\mu$, $m_\tau$, and 
the masses of all other particles from units $m_\circ$ into 
proton mass units, $m_p$, thus enabling these masses to be compared 
with the experimental data. 
The computed fermion masses are listed  
in Table \ref{t:massresult} where the symbols  ${\Y}_1$, ${\Y}_2$ and
${\Y}_3$ denote complex ``helical'' tripoles that
replicate the properties of the simple tripole {\Y} on higher
levels of the hierarchy.
These helical tripoles can be regarded as the combinations 
of ``heavy neutrinos'' with simple triplets. Like $\nu_e$, 
the heavy neutrino consists of six pairs of helical triplets:
$\nu_{\rm h} =6{\Y}_1\overline{{\Y}}_1.$ 
They can further combine and form ``ultra-heavy'' neutrinos 
$\nu_{\rm \rum uh} =3\rub (\overline{\Y}_1 \nu_{\rm \, h} u)e^-$ and so on.
The components ${{\Y}_2}$ and ${{\Y}_3}$ of  
the $c$ and $t$ quarks have the following structures: 
${{\Y}_2}\,{=}$\linebreak ${=}\,u\nu_e u\nu_e e^-$, 
consisting of 165 primitive particles,
and ${{\Y}_3}\,{=}$ ${=}\, \nu_{\rm uh}${\Y}, 
consisting of 1767 primitive particles.

\markright{V.\,N.\,Yershov. Fermions as Topological Objects}

\section{Conclusions}

\markright{V.\,N.\,Yershov. Fermions as Topological Objects}

The results presented in Table~\ref{t:massresult} show that 
our model agree with experiment to an accuracy better then 0.5\%.
The dis\-crep\-ancies should be attributed to the simplifications we have
assumed here
(e.\,g., neglecting the binding and oscillatory energies, as well as
the neutrino residual masses, which contribute to the masses of many structures
in our model).

By matching the pattern of properties of the 
fundamental particles our results confirm that 
our conjecture about the dualism of space and the symmetry of the
basic field cor\-res\-ponds, by a grand degree of 
confidence, to the actual situation. Thus, our model   
seems to unravel a new layer of physical reality, which bears 
the causal mechanisms under\-ly\-ing quantum phenomena. This 
sets a foundation from which one can explain many otherwise inexplicable
observational facts that plague modern physics.


\markright{V.\,N.\,Yershov. Fermions as Topological Objects}

\section*{Acknowledgements}

\markright{V.\,N.\,Yershov. Fermions as Topological Objects}

The author thanks Prof. V.\,V.\,Orlov for his valuable com\-ments
and Dr. L.\,V.\,Morrison for his linguistic
support.

}
\centerline{\rule{0pt}{10pt}\rule{72pt}{0.4pt}}

\end{document}